**A new approach to simulation of limiting photoconversion efficiency of tandem solar cells**

A.V. Sachenko[1], V.P. Kostylyov[1], N.P. Kulish[1], I.O. Sokolovskyi[1] and A.I. Shkrebtii[2]

[1]V.Lashkaryov Institute of Semiconductor Physics, NAS, Kyiv, Ukraine
[2]University of Ontario Institute of Technology, Oshawa, ON, Canada

*Corresponding author e-mail: sach@isp.kiev.ua and anatoli.chkrebtii@uoit.ca*

**Abstract**

We develop a new approach to calculate the obtainable limit of photoconversion efficiency $\eta$ of tandem solar cells (SCs) and applied it to SCs with both vertical and lateral designs at AM0 and AM1.5 conditions. To get the maximum efficiency, only radiative recombination has been considered using typical radiative recombination parameters of the direct band gap III–V semiconductors, and explicit energy dependence of light absorption. When simulating the efficiency, we selfconsistently took into account the fact that the amount of the heat dissipated by SC decreases as the number of current-matched subcells $n$ increases. As the operating SCs temperature decreases both the open-circuit voltage $V_s$ and the photoconversion efficiency $\eta$ increase. It is shown that the above effect is especially strong for SCs operating under AM0 conditions. As the number of subcells $n$ is increased, narrowing the spectral range for each subcell, the photocurrent is additionally reduced due to the energy dependent light absorption, the factor generally ignored in the standard approaches. Application of our formalism results in a maximum in the theoretical $\eta(n)$ dependence, which was indeed observed experimentally. Besides agreement with experiment, our theoretical results are also close to other efficiencies calculated using detailed balance based approaches.

## 1. INTRODUCTION

Multi-junction or tandem solar cells offer a realistic way to essentially increase efficiency of solar-energy conversion into electrical power for both non-concentrated and concentrated solar radiation (see, *e.g*., [1-3] and references therein). Since this should be complemented with price reduction of the produced electricity, optimization of the solar cells in terms of the parameters and their design is very important both fundamentally and technologically. To predict a realistic attainable efficiency of the solar-energy conversion, one should account for the fundamental constraints that restrict the photoconversion processes thus setting targets to both research labs and industry. Simulations of maximum attainable photoconversion efficiency (from now on efficiency) of a single-junction solar cell (SC), assuming that all the recombination channels, ex-cept radiative one, can be neglected, was performed by Shockley and Queisser in their classic work [4] (extended discussion of the detailed balance approach can be found in [5]). Later several approaches for calculating the efficiency limit of tandem (*i.e.*, multijunction) SCs, operating under non-concentrated and concentrated irradiation, were proposed in [2,6,7,9-10] and references therein. The absolute upper efficiency maximum for photovoltaics is the Carnot limit and the efficiency values obtained using thermodynamics based approaches by Landsberg-Tonge [7] and de Vos-Grosjean-Pauwels [9] are close to that limit. However, achieving this limiting Carnot efficiency is nearly impossible, even in principle. More realistic calculations of the efficiency are offered by detailed balance approach, with some certain assumptions for photoconversion processes, such as in [4], by considering black body radiation from the Sun surface and only radiative recombination, which sets an upper limit to the minority carrier lifetime. Detailed balance approach applied to the tandem SCs (see, *e.g.*, [1]) demonstrated monotonic efficiency increase with the number of subcells $n$. Fig. 4 from [2] summarizes efficiency limits obtained with four different approaches depending on the ratio between the temperatures of radiation source and radiation detector. The lowest (and closest to the experimentally observed) efficiency is derived in the Shockley-Queisser approach [4]. The efficiency limits of tandem SCs from the de Vos-Grosjean-Pauwels and Shockley-Queisser approaches demonstrate that the formalism of [4] offers more realistic efficiencies as compared to the experimental values. Authors of [6] offered an approach to calculate the efficiency of tandem SC at AM0 that considers both radiative and Shockley-Read recombinations in the space-charge-region. They calculated $\eta$ for a set of temperatures 200, 300, 400 and 500 K, considering radiative recombination only.

However, the approaches of [2,5,12] do not account for another fundamental constraint for the tandem SC efficiency, namely the degree of light absorption close to the fundamental absorption edge of each subcell. At this edge the external quantum efficiency (EQE) (or photocurrent quantum yield) $q_s(E_{ph})$ strongly depends on the photon energy $E_{ph}$ and is below one. However, the ideal quantum efficiency with a square shape and $q_s = 1$ across the entire spectrum of wavelengths is usually assumed for each subcell [2,12]. Apart from the above band gap factor, solar cell optimization should selfconsistently include the natural effect of SC heating up. Therefore, it is necessary to develop new simulation tools to study tandem SC efficiency limits considering the above mentioned photoconversion constraints in real SCs. The goal of the present work is to implement those factors into photoconversion simulation formalism.

Our key assumption (as in [4] and other similar approaches) is that radiative recombination is the only recombination mechanism. In our calculations we used the radiative recombination parameters typical for the direct band gap III−V semiconductors. In the framework of equiv-alent circuit we considered series-interconnected current-matched subcells. In addition, our formalism accounts for the temperature balance, that is overall SCs temperature decrease as the number of cells *n* increases. The reason is that for a large number of subcells, the difference between the band gaps and photon energies decreases and less solar energy is transformed into heat caused by the effect of photoexcited carriers' thermalization. Such temperature reduction leads to the increase of the open circuit voltage and, subsequently, efficiency. This effect is strongest for SCs operating in outer space, where only radiation mechanisms of SC cooling are present and the lowest SC temperature is not limited by its reasonably high value under terrestrial conditions.

Important is that we also accounted for another fundamental effect that reduces the limiting value of the photocurrent as the number of cells *n* increases, namely the energy dependent light absorption. The photocurrent quantum yield $q_s(E_{ph})$ as a function of photon energy $E_{ph}$ near the absorption edge is below its maximal value of one. Therefore, if the solar spectrum is distributed over a larger number of subcells, the area below the curve $q_s(E_{ph})$, which is equal to the photocurrent in the photon energy range $\Delta E_{ph} = E_{ph2} - E_{ph1}$ for each subcell, declines more rapidly than the photon flux. Subsequently, despite the photovoltage increase with *n* due to a larger number of energy intervals, the photocurrent starts decreasing due to lower quantum yield, thus causing the efficiency $\eta$ to decrease with the number of subcells and leading to a maximum in the $\eta(n)$ curve. Our results, when incorporating these two effects into the formalism, are in good agreement with experiments and improve existing theoretical approaches (see, e.g., [1] and refs therein) that result in a continuous gradual increase of the efficiency $\eta(n)$.

## 2. FORMULATION OF THE PROBLEM

The simulation of obtainable efficiency of tandem SCs was performed self-consistently by solving equations for photocurrent, photovoltage and thermal balance for both AM0 and AM1.5 conditions. We considered two present-day designs of tandem SCs, namely, vertical [13] and lateral [14]. In the first design, the $p-n$ junctions or heterojunctions with different band gaps are placed on top of each other, with wider gaps on the top and the gap narrowing through the body of the cell. In the lateral design, all the $p-n$ junctions (or heterojunctions) are in the same plane, with only a part of solar spectrum falling on each of them. Spectrum splitting is made with a dispersion element (*e.g.*, diffraction grating), with the splitting efficiency close to 100%.

As noted above, the temperature of the tandem SC surface decreases with $n$ as the energy gap interval for each subcell in the tandem becomes narrower. If the thermal conductivity of materials of single cells is high enough (so that the same stationary temperature is set), then the absolute temperature of SC can be found from the thermal balance equation (see, for instance, [15]):

$$\int_{E_1}^{E_2} P(E_{ph})dE_{ph} - J_m(T)V_m(T) = \beta\sigma T^4 . \qquad (1)$$

Here $P(E_{ph})$ is specific power of solar radiation at a given photon energy $E_{ph}$, $E_1$ and $E_2$ are the lower and upper photon energy limits for the entire SC, $J_m(T)$ and $V_m(T)$ are the photocurrent density and photovoltage at maximum power output. The parameter $\beta$ is of the order of unity and defines a degree to which the SC radiates, compared to perfect black body emission, and the extra contributions to the heat up depending on the vicinity of the SC to the satellite and its orientation. Parameter $\sigma$ is the Stefan-Boltzmann constant, and $T = T_{\min} + \Delta T$, with $T_{\min}$ being the ambient temperature. From here on, we suppose that the surface of a tandem SC under consideration is of unit area.

Let us analyse in more detail the minimum temperatures of solar batteries (SBs) under AM1.5 and AM0 conditions. Under AM1.5 the minimum possible SC temperature $T_{min}$ is limited by the ambient temperature and we assume for the sake of simplicity that the only cooling mechanism is due to radiation. Under AM0 the SC heats up due to partial conversion of solar energy into heat inside the cell as well as absorption due to both sunlight reflected from the satellite and its own heat emission. It is possible, however, to avoid the last effect by properly orienting the SC and the satellite. Friction with residual atmosphere can also cause extra heating up, but this mechanism depends on the satellite orbit and orientation of the SC relative to the satellite direction of motion. For example, $T_{\min}$ is higher for low altitude satellite orbits. It was shown in [16], however, that even in the case of low orbits $T_{\min}$ due to friction is about 170K. This is considerably smaller than the temperature of SC due to radiation and in what follows, the frictional contribution to the total SC temperature under AM0 will be neglected.

In our calculations we considered that SCs absorb solar radiation in the wavelength range of $0.3\mu m < \lambda < 2\mu m$ (or 4.13 ÷ 0.62 eV energy range): the low and high energy parts of the solar spectrum are cut off by the top transparent layer of a tandem solar cell or optical splitter for lateral SC. Therefore, non-photoactive low energy photons with $\lambda > 2\mu m$ do not contribute to SCs temperature. Fig. 1, calculated according to Eq. (1), shows temperature $T$ of a SC vs. its efficien-cy $\eta$ under AM0 condition. At $\beta = 2$ SC emits as a black body, at $\beta = 1$ the cell radiates as a grey body with emission coefficient of 0.5, while $\beta=1.5$ corresponds to the intermediate case. Two sets of curves are given for each $\beta$: the upper (solid) curve corresponds to the temperature from the entire solar spectrum, while the lower curve (dashes) is calculated without the contribution of the low-energy part of the spectrum ($\lambda > 2\mu m$ or $E_0 \leq 0.62$ eV), as indicated by arrows. Cutting off the low energy part of the solar spectrum further reduces SCs temperature.

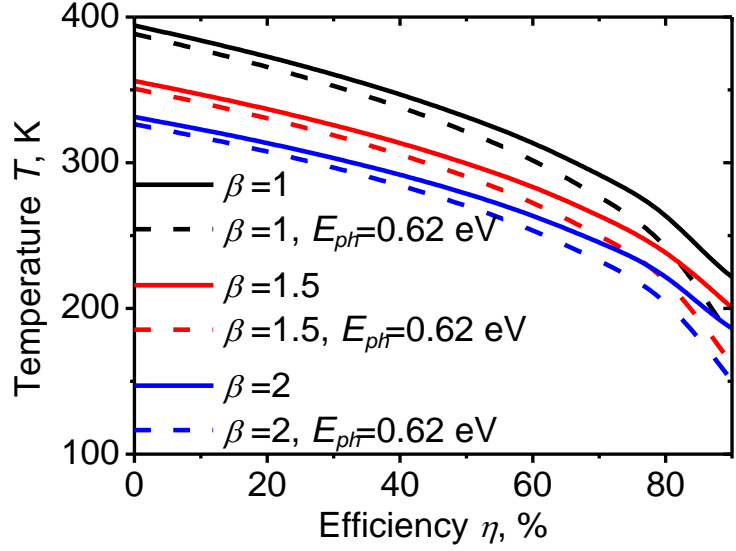

Fig. 1. (Color online) Calculated temperature dependent efficiency $\eta(T)$ for AM0 conditions of SCs at different cooling rates at $\beta = 1$ (solid and dashed lines on top), $\beta = 2$ (solid and dashed line at the bottom) and intermediate case $\beta = 1.5$.

To derive the expressions for photovoltage and photocurrent for series-interconnected current-matched subcells at maximum power output, we neglect the series resistance $R_s$ of the tandem cell stacks and consider that the strong inequality $L_i \gg d_i$ always holds ($L_i$ is the diffusion length of minority carriers and $d_i$ is the $i$-th subcell thickness). If under open circuit condition the excess concentration of minority charge carriers in the $i$-th subcell $\Delta n_i$ is considerably smaller than the equilibrium concentration of majority carriers $n_{0i}$, the open circuit electromotive force (emf) $V_i$ for the $i$-th cell is given by the well-known expression [17]:

$$V_i = \frac{kT}{q} \ln\left(\frac{\Delta n_i n_{0i}}{n_i^2(T)}\right). \qquad (2)$$

Here $k$ is the Boltzmann constant, $q$ is the elementary charge, and the concentration of intrinsic charge carriers $n_i(T)$ is defined in terms of the effective density of states in the conduction (valence) band $N_{ci}$ ($N_{vi}$) and the band gap $E_{gi}$ for the $i$-th subcell, and at $T = 300$ K

$$n_i(T) = \sqrt{N_{ci} N_{vi}} \left(\frac{T}{300}\right)^{3/2} \exp\left(-\frac{qE_{gi}}{2kT}\right). \tag{3}$$

Using the above approximations for the minority carriers, their excess concentration $\Delta n_i$ can be found from the generation-recombination balance equation

$$J_{gi} = q\Delta n_i d_i A_i n_{0i}. \tag{4}$$

Here $J_{gi}$ is the short-circuit current density in the $i$-th cell, and $A_i$ is the radiative recombination constant for the $i$-th semiconductor. Substitution of Eqs. (3) and (4) into Eq. (2) gives the following final expression for $V_i$:

$$V_i = E_{gi} - \frac{kT}{q}\ln(1+v_i), \tag{5}$$

where

$$v_i = \frac{qd_i A_i N_{ci} N_{vi} \left(\dfrac{T}{300}\right)^3}{J_{gi}}. \tag{6}$$

In the case of near-unity ideality factor of the $I$–$V$ curve, the expression for $V_i$ can be simplified as:

$$V_i = \frac{kT}{q}\ln\left(\frac{J_{gi}}{J_{0i}} + 1\right), \tag{7}$$

where $J_{0i}$ is the saturation current density for the $i$-th subcell. For $n$ interconnected in series current-matched cells we can write the following expression for the open-circuit voltage $V_S$ [18]:

$$V_S = \frac{kT}{q}\sum_i^n \ln\left(\frac{J_{gi}}{J_{0i}} + 1\right) = n\frac{kT}{q}\ln\left(\frac{J_{gi}}{J_{0s}} + 1\right), \tag{8}$$

where $J_{0s} = (J_{01} \cdot J_{02} \cdot \ldots \cdot J_{0n})^{1/n}$.

Using Eqs. (2) – (8), the expression for $V_S$ can be simplified to:

$$V_S = \sum_i^n E_{gi} - \frac{kT}{q}\sum_i^n \ln(1+v_i). \tag{9}$$

Finally, if all $v_i$ are close such the $\ln(1+v_i)$ can be considered the same for each subcell (see a discussion about this later), Eq. (9) takes the following form:

$$V_s = \sum_i^n E_{gi} - \frac{kT}{q} n \ln(1+v). \qquad (10)$$

One sees from Eq. (9) or (10) that in two limiting cases (i) when $T = 0$, or (ii) in the absence of recombination with $A_i = 0$, the open-circuit voltage of tandem SC is maximal and simply equals to the sum of band gaps of the semiconductors used. It is important that in real devices the strong inequality $v_i \gg 1$ holds and in this case the open circuit voltage $V_S$ from Eq. (9) or (10) decreases linearly with the temperature $T$.

When a load is connected to SC, the current density $J(V)$, that is current-voltage characteristic or $I$–$V$ curve, can be found following [17]:

$$J(V) = J_g - J_{0s} \exp\left(\frac{qV}{nkT}\right), \qquad (11)$$

where $V$ is the SC voltage, $J_g$ is the SC photocurrent density and $J_{0s}$ its saturation current density. To a good approximation the current $J_m$ and voltage $V_m$ that yield the maximum SC power output $P_m$ ($P_m = J_m \cdot V_m$), can be expressed as:

$$J_m \cong J_g\left(1 - \frac{nkT}{qV_s}\right), \qquad (12)$$

$$V_m \cong V_s\left(1 - \frac{nkT \ln(qV_s/nkT)}{qV_s}\right). \qquad (13)$$

Eqs. (12) and (13) have to be used to calculate the efficiency $\eta$ of tandem SC:

$$\eta = J_m V_m / P_s, \qquad (14)$$

where $P_s$ is the incoming solar power the SC is exposed to. Since the SC temperature $T$ enters in both equations (12) for $J_m$ and (13) for $V_m$, $T$ should be selfconsistently included in the optimization formalism through the thermal balance equation. The set of Eqs. (1)–(14) have to be applied to optimize both lateral and vertical tandem SCs under consideration.

To calculate the efficiency $\eta$ we begin with the short-circuit current $J_m$ for lateral SC. First, it is important to stress that contrary to the commonly considered approximation (see, *e.g.*, [12]) with a photocurrent quantum yield of unity, we include in the formalism an actual energy dependence of the yield, which depends on the absorption of light and for each subcell *always starts from zero at the band edge*, increasing to one with photon energy. Such energy dependence of photocurrent quantum yield $q_{si}(E_{gi}, E_{ph})$ causes an incomplete light absorption in semiconductor close to the band edge and leads to the fundamental limitation of the photoconversion efficiency for a large number of subcells. Taking this into account the following expression for the short-circuit current densities of the single $i$-th cell $J_{gi}^L$ can be written:

$$J_{gi}^L(E_{gi}) = s_i^{-1} \int_{E_1}^{E_2} j_{gi}(E_{ph}) q_{si}(E_{gi}, E_{ph}) dE_{ph}. \tag{15}$$

Here parameter $s_i$ is the ratio between the surface area of the $i$-th subcell and the total surface area of the lateral SC, $j_{gi}(E_{ph})$ is the photocurrent density at photon energy $E_{ph}$, $j_g(E_{ph}) = q \cdot P(E_{ph})/E_{ph}$ with electron charge $q$ and solar radiation flux density $P(E_{ph})$. A general expression for $q_{si}(E_{gi}, E_{ph})$ was discussed in [19,20]. In the limit when the base width is significantly larger than the diffusion length, i.e., $d_i \gg L_i$, the emitter width is negligible compared to $d_i$, the recombination velocities at the front and back surfaces of the $i$-th cell are close to zero, and the expression (15) is considerably simplified:

$$q_{si}(E_{gi}, E_{ph}) = \frac{\alpha_i(E_{gi}, E_{ph}) L_i}{1 - (\alpha_i(E_{gi}, E_{ph}) L_i)^2} \left[ \frac{d_i}{L_i} + \alpha_i(E_{gi}, E_{ph}) L_i \exp(-\alpha_i(E_{gi}, E_{ph}) d_i) - \alpha_i(E_{g_i}, E_{ph}) L_i \right]$$

(16)

where $\alpha(E_{gi}, E_{ph})$ is the light absorption coefficient, a collection of experimental absorption coefficients can be found, e.g., in "New Semiconductor Materials Database" [21].

To calculate how the efficiency $\eta$ of the lateral tandem SC depends on the number of subcells $n$, the system of equations (1) and (10) – (16) is to be solved selfconsistently with the photocurrent-matching condition satisfied. The current matching implies that the photocurrent of each subcell is the same, i.e.:

$$J_{g1}(E_{g1}) = J_{g2}(E_{g2}) = \cdots = J_{gn}(E_{gn}). \tag{17}$$

Here for the lateral SC we consider $J_{gi}^L(E_{gi}) = J_{gi}(E_{gi})$ and $J_{gi}^V(E_{gi}) = J_{gi}(E_{gi})$ for the vertical SC.

For the vertical tandem SCs, the short-circuit currents of different subcells $J_{gi}^V$ is calculated by taking into account that solar light goes through a stack of single subcells, and:

$$J_{gi}^V(E_{gi}) = \int_{E_1}^{E_2} j_g(E_{ph}) q_{si}(E_{gi}, E_{ph}) T_i(E_{gi}, E_{ph}) dE_{ph}, \tag{18}$$

where

$$q_{si}(E_{gi}, E_{ph}) = \frac{\alpha_i L_i}{1 - (\alpha_i L_i)^2} \left[ \frac{d_i}{L_i} + \alpha_i L_i \exp(-\alpha_i d_i) - \alpha_i L_i \right], \tag{19}$$

$$T_i(E_{gi}, E_{ph}) = e^{-\alpha_1 d_1 - \alpha_2 d_2 - \cdots - \alpha_i d_{i-1}}. \qquad (20)$$

The coefficients $e^{-\alpha_i d_i}$ account for the solar light intensity decrease when light passes through the $i$-th cell. In this case, to determine the efficiency $\eta$ as a function of number of cells $n$ Eqs. (1), (10) – (14) and (18) – (20) are solved selfconsistently. As for the case of lateral SC, it is necessary to additionally satisfy the photocurrent matching condition, as defined by Eq. (18). In the following section we demonstrate how the efficiency depends on the temperature and number of the subcells at different conditions.

## 3. THE MAIN RESULTS AND ANALYSIS

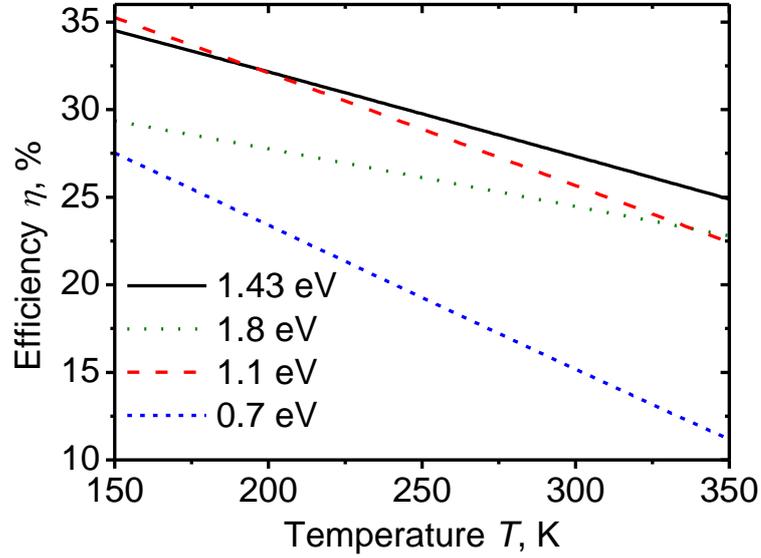

Fig. 2. (Color online) Calculated temperature dependent efficiency $\eta(T)$ at AM0 for single-junction SCs with the band gaps $E_g$ equal to 0.7 eV (short dashes), 1.1 eV (dashes), 1.43 eV (solid line) and 1.8 eV (dots).

Fig. 2 demonstrates the temperature dependences of efficiency $\eta(T)$ for four different single-junction SCs with band gaps of 1.8, 1.43, 1.1 and 0.7 eV under AM0 conditions and incoming solar radiation in the 0.3–2 μm wavelength range for temperatures from 150 K to 350 K. The efficiency $\eta$ decreases with $T$ at a higher rate for narrow-gap semiconductors, while $\eta$ demonstrates a slower decrease for wide-gap semiconductors. For example, $\eta$ for GaAs at $E_g$=1.43 eV is less than that for $E_g$=1.1 eV in the temperature range from 150 to 200 K, while this inequality reverses above 200 K. Such temperature dependent efficiency correlates with variations of the open-circuit voltage of the SC. Obviously, enhanced efficiency can be achieved by forced cool-ing. However, Fig. 2 offers understanding of the efficiency variation trends for tandem SCs operating under AM0 conditions with increasing subcells number $n$, which we discuss next.

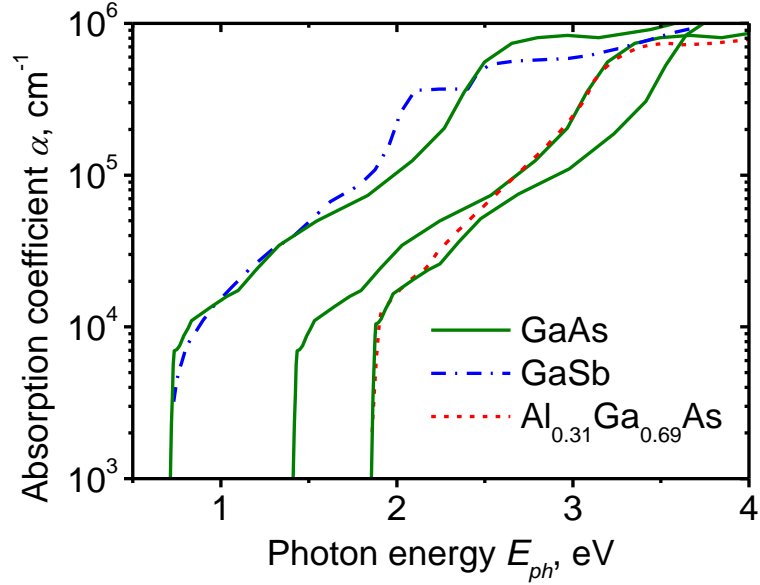

Fig. 3. (Color online) Experimental light absorption coefficient *vs.* photon energy $\alpha(E_g, E_{ph})$ for the set of direct gap semiconductors (from [21]).

Fig. 3 is used to demonstrate how the absorption coefficient dependence on the photon energy $E_{ph}$ for arbitrary gap $E_g$, namely $\alpha(E_g, E_{ph})$, that is required for calculations, can be obtained. As an example, we plotted in Fig. 3 the experimental $\alpha(E_g, E_{ph})$ dependences, taken from [21], for three direct-gap reference semiconductors GaSb (dot-dashes), GaAs (solid line) and $Al_{0.31}Ga_{0.69}As$ (dashes). For instance, a rigid shift of the experimental GaAs $\alpha(E_g, E_{ph})$ curve towards lower energy by the difference in $E_g$ with GaSb, demonstrates that the experimental $\alpha(E_g, E_{ph})$ dependence for GaSb to a good accuracy can be matched by the absorption curve for GaAs. Similar shift of the $\alpha(E_g, E_{ph})$ curve for GaAs towards higher $E$ matches it with a good accuracy the $Al_{0.31}Ga_{0.69}As$ curve. In both cases, a good matching is achieved in the region where $\alpha(E_g, E_{ph})d_i \leq 1$. The fact that there is no good match when $\alpha(E_g, E_{ph})d_i \gg 1$, is not important since the photocurrent quantum yield in this case is unity and light is completely absorbed. This procedure allows us to rigidly shift the experimental $\alpha(E_g, E_{ph})$ curve for GaAs to extrapolate the absorption coefficient of another semiconductor with arbitrary $E_g$ without compromising the accuracy of the calculated $\eta$.

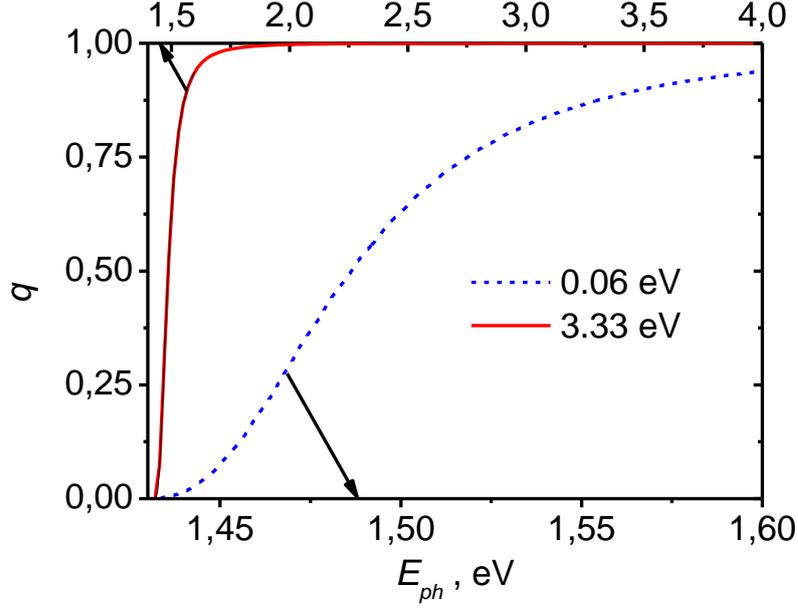

Fig. 4. (Color online) Calculated photocurrent quantum yield *vs.* photon energy plotted over a wide (top horizontal axis, solid line) and magnified narrow (bottom horizontal axis, dotted line) energy intervals.

In the next section we analyze the change in photocurrent quantum yield $q_s$ change with increasing number of subcells *n*, leading to narrowing the energy range $\Delta E_{ph} = (E_1 - E_2)$ for *i*-th subcell (we consider here that $E_1 = E_{gi}$). Fig. 4 (solid line) shows how $q_s(E_g, E_{ph})$ depends on the photon energy $E_{ph}$ for a semiconductor with the band gap $E_g$=0.67 eV across the energy interval of 3.33 eV (top axis), as calculated using Eq. (16). To magnify the steep variation at low $E_{ph}$, the curve is re-plotted in a small energy interval of 0.06 eV (bottom axis). One can see from Fig. 4 that, for a wide energy range, the area under the curve $q_s(E_g, E_{ph})$ is close to that of a rectangle of unit height. For the case of narrow energy range, the corresponding area (that characterizes photocurrent) is considerably less than that of the above rectangle. Therefore, as the energy interval $\Delta E_{ph}$ narrows down, the photocurrent decreases more rapidly than the photon flux in this range, resulting in additional decrease of the photocurrent. With increasing number of subcells *n* the above effect will compete with effect of $V_m(T)$ increase (which also reduces SC temperature) and will eventually result in appearance of the maximum on the $\eta(n)$ dependence.

To calculate how the efficiency $\eta$ depends on *n* for a lateral tandem SC under AM0 and AM1.5 conditions assuming the availability of semiconductors with any required $E_g$ to satisfy the current matching, we further simplify the set of the main equations. Substituting the parameters of direct-gap III–V semiconductors from [21] for a reference SCs into Eq. (9), we find that $v_i \gg 1$ and $v_i$ values are such that $\ln(1+v_i)$ are close enough for all the semiconductors considered. This allows replacing Eq. (9) by Eq. (10) and using in Eq. (10) only parameters for GaAs, namely $A = 2 \times 10^{-10}$ cm$^3$/s, $D = (10 \div 50)$ cm$^3$/s, $N_{c0} = 5 \times 10^{17}$ cm$^{-3}$ and $N_{v0} = 1 \times 10^{19}$ cm$^{-3}$. We also assume that subcells' width $d = d_i = 2 \times 10^{-4}$ cm and doping level $n_0 = n_{0i} = 10^{17}$ cm$^{-3}$ are the same. The above parameters, together with the extrapolated $\alpha(E_g, E_{ph})$ dependencies, $v$ values and $E_g$, required to satisfy the current matching conditions, were used to calculate $\eta(n)$.

Fig. 5 (a) and (b) combine the calculated attainable efficiency $\eta$ (the left axis) and $T$ (right axis) as a function of $n$ for lateral tandem SCs. The figures correspond to AM0 and AM1.5 conditions respectively. We supposed that all the subcell widths along the lateral SC surface are the same and their sum equals to the distance where the dispersion element decomposes the solar light. For the tandem SCs under AM0 conditions and $\beta = 2$ (SC radiates as a perfect black body) the maximum attainable efficiency $\eta(n) = 54.4\%$ is achieved when 15 subcells are used. Under AM1.5 and $\beta=2$, $\eta(n)$ reaches its maximum of 49.6% at $n \approx 12$ (see Fig. 5b). We have to stress that the commonly used approaches (see, *e.g.*, [12]) do not demonstrate a maximum of the $\eta(n)$ curve.

Using Fig. 5, we correlate $\eta$ (right axis) and $T$ (left axis, calculated from Eq. (1)) both as a function of $n$. For both AM0 and AM1.5 conditions the lower is SC temperature the higher the efficiency. Under AM1.5 conditions, the SC temperature is always higher than the ambient temperature $T_{\min} = 300$ K. Under AM0 and at sufficiently large $n$, the SC temperature decreases and saturates at 264 K at $\beta = 2$ and 287 K at $\beta = 1.5$. One can also see from Fig. 5a that at $\beta = 2$ (SC radiates as a perfect black body) its temperature is close to 300 K, even for a single junction solar cell ($n = 1$). It should be noted that for SCs at $n = 1$, operating under AM0 the actual $\beta$ value is close to 1.5.

We assumed in Figs. 5 (a) and (b) that each subcell is made of a direct-gap III–V semiconductor with the $E_{gi}$ values satisfying current matching; we label this hypothetical system the model case. In principle, ternary $A_xB_{1-x}C$ semiconductors allow producing a required set of semiconductors with continuous band gaps [13]. In reality, however, both binary and ternary semiconductors with $E_g \geq 2$ eV are indirect and a thicker SC with $d$ (~100 μm) is required as a result, which creates technological challenges when producing lateral tandem SCs. Therefore, it is important to model $\eta(n)$ using parameters of existing direct-gap semiconductors. In this case, a discrepancy between available and required $E_g$ together with the need to match the current in different subcells further reduces efficiency, and $\eta(n)$ reaches its maximum at smaller $n$ as compared to the model case shown in Fig. 5.

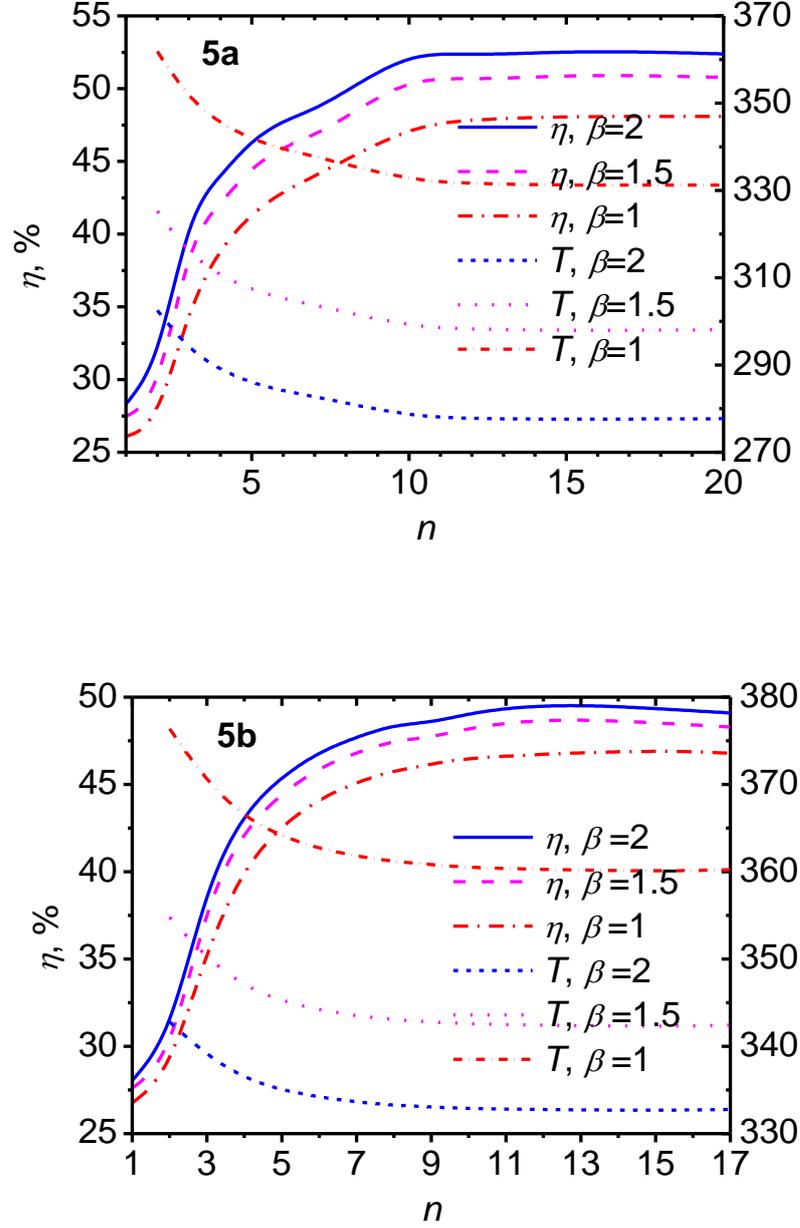

Fig. 5. (Color online) Calculated solar cells efficiency $\eta$ (left axis) and temperature $T$ (right axis) *vs.* number $n$ of *p-n*-junctions for lateral SC under AM0 (5a) and AM1.5 (5b) conditions and different cooling ($\beta = 1$, $\beta = 1.5$ and $\beta = 2$).

The simulated $\eta(n)$ for such realistic SCs under different operational conditions are presented in Fig. 6. They demonstrate significantly lower efficiencies as compared to the model case. The pronounced maximum of $\eta(n)$ is, however, reached at *only four subcells* under AM0 conditions and *five subcells* under AM1.5. While the efficiency $\eta(n)$ will depend on the actual set of semiconductors used in the concrete tandem SC, the qualitative picture will remain the same. As in the previous model case, the solar cell temperature is lower and the efficiency is higher at $\beta = 2$ when the SC radiates as a perfect black body compared to cases where $\beta$ is smaller. Under AM0 conditions and $\beta = 2$ the maximum efficiency $\eta_{max} = 45.6\%$ is reached at $n = 4$. (This is considerably less than $\eta = 54.4\%$ at n=15 for the model case, shown in Fig. 5). For a comparison, Fig. 6 also shows experimental $\eta(n)$ plot (as compiled from [14,22,23]), that demonstrates reasonably good agreement with our theoretical results, including the presence of the maximum.

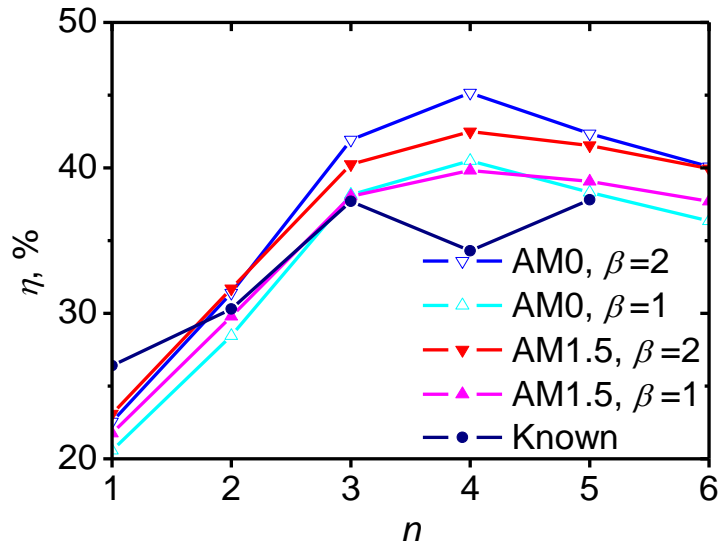

Fig. 6. (Color online) Calculated efficiency $\eta$ *vs.* number of subcells $n$ (for a set of experimentally available semiconductors) for AM0 and AM1.5 conditions and different cooling ($\beta = 1$, $\beta = 1.5$ and $\beta = 2$) for the lateral solar cells. Calculated efficiencies are compared with experimentally available efficiencies (from [14, 22, 23]).

Our developed approach is also valid for SCs under concentrated radiation. The mechanism of the increase in $\eta(n)$ under AM0 and light concentration due to a decrease in SC temperature is similar to the case of non-concentrated solar radiation. In both cases $\eta$ increases due to an increase of open-circuit voltage $V_s$. A comparison of the temperature induced efficiency variation shows that the increase of $\eta(n)$ when $T$ decreases to 261 K from 303 K is similar to the efficiency increase that caused by tenfold concentration of solar radiation. It should be noted that in all the cases considered, the calculated $\eta(n)$ under AM0 conditions exceed the efficiencies under AM1.5.

Let us analyze the SCs efficiency $\eta(n)$ under concentrated radiation as a function of the degree of sunlight concentration $M$ in the outer space environment. Since under space conditions SC cooling is only due to heat emission, the only way to offset the temperature increase due to increase of $M$ is to increase the heat emitting surface by substantial increase of radiators area. Otherwise a considerable SC temperature increase will cause a large efficiency drop. In the particular case when the area of radiators is proportional to $M$, the dependence on $M$ cancels in all the three terms in the thermal balance Eq. (1). However, the $M$ dependency still enters the second term of Eq. (9) or (10) for $V_m$ because $J_g(M) = J_g(1) \cdot M$, which results in an increase of the open-circuit voltage $V_s$ and the efficiency $\eta(n)$ with $M$.

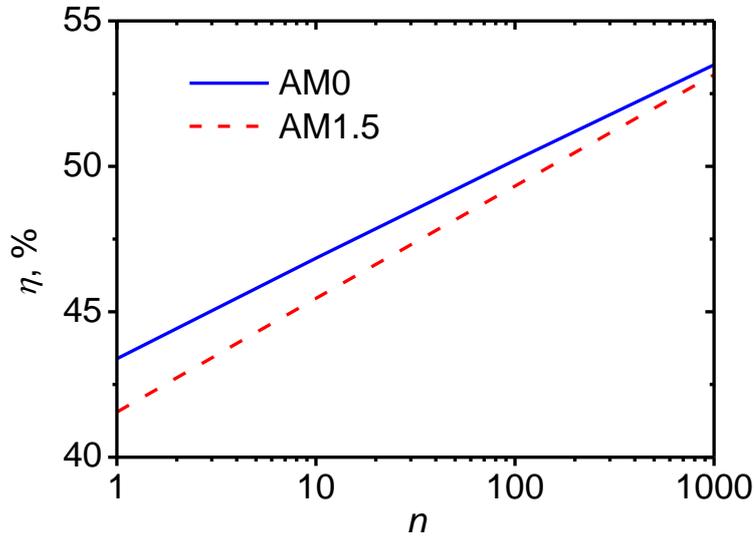

Fig. 7. (Color online) Calculated efficiencies $\eta$ vs. degree of solar light concentration $M$ for AM0 and AM1.5 conditions.

Fig. 7 presents the efficiency as a function of the degree of concentration $M$ under AM0 (upper line) for four-junction SC. Its efficiency for non-concentrated light ($M=1$) is 43.4% and at $M = 10^3$ $\eta = 53.5\%$, and representing about half a percent of efficiency increase due to a small decrease of the SC temperature. Calculated in a similar manner as for the AM0 case, the lower line in Fig. 7 demonstrates $\eta(M)$ for terrestrial conditions at AM1.5. Although in this case the minimal SC temperature is limited by the ambient temperature, if the radiation cooling mechanism is predominant for AM1.5, $\eta(M)$ dependence is similar to that one for the space conditions.

However, since at AM1.5 the SC temperature is higher, the efficiency $\eta$ is smaller than that one at AM0 conditions.

All the calculations of $\eta(n)$ were made by neglecting the effect of series SC resistance. Since this is not the case for concentrated sunlight, one has to expect a higher discrepancy between the calculated and experimental results for $\eta(n)$.

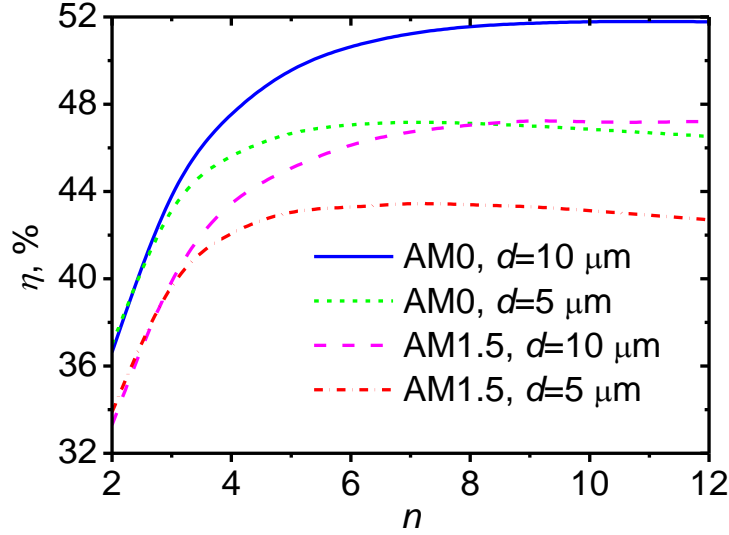

Fig. 8. (Color online) Calculated $\eta$ dependencies for different cells thicknesses *vs.* number *n* of *p-n*-junctions for AM0 and AM1.5 conditions (vertical solar cells, hypothetical model case for required energy gaps).

Fig. 8 shows the calculated efficiency $\eta(n)$ for the model vertical tandem SCs operating under AM0 and AM1.5 conditions at $\beta = 2$. We considered two 5 μm and 10 μm thick SCs (this is the total thickness of all subcells composed) with all other parameters being the same as before and solved selfconsistently Eq. (1)–(14) and (18)–(21). For the same thickness $\eta$ under AM0 is higher than for AM1.5. At $d = 5$ μm $\eta$ is lower than for $d = 10$ μm, and it reaches maximum at smaller number of subcells $n = 7$ at $d = 5$ μm. At $d = 10$ μm the maximum efficiency is comparable to that of lateral structures. Lower efficiency for the $d = 5$ μm thick SC is due to the higher losses related to the incomplete light absorption in each subcell if their number increases considerably. Therefore, for thinner SC not only the maximum of $\eta(n)$, but the number of cells, at which this maximum efficiency is achieved, is decreased.

The calculated efficiency $\eta(n)$ under AM1.5 conditions for model systems with lateral and vertical structures are shown in Fig. 9. In the same figure we added the theoretical $\eta(n)$ from [1]. The overall agreement between the results of our and other detailed balance approaches [1], which were applied to the tandem SCs, is reasonably good. All the above efficiencies, however, are considerably smaller than those calculated by Landsberg-Tonge and de Vos-Grosjean-Pauwels [7-9] using thermodynamics approaches, which considerably exceed experimentally available efficiencies.

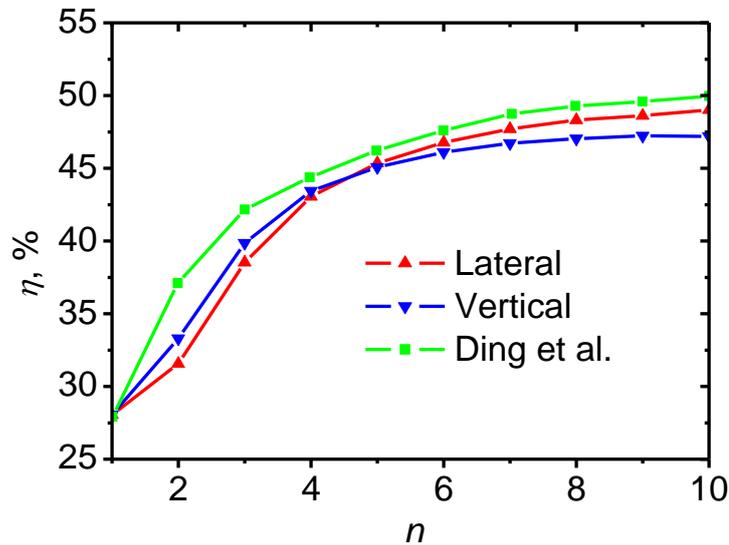

Fig. 9. (Color online) Calculated efficiency $\eta$ versus number of cells for model systems with lateral and vertical structures for AM1.5 conditions.

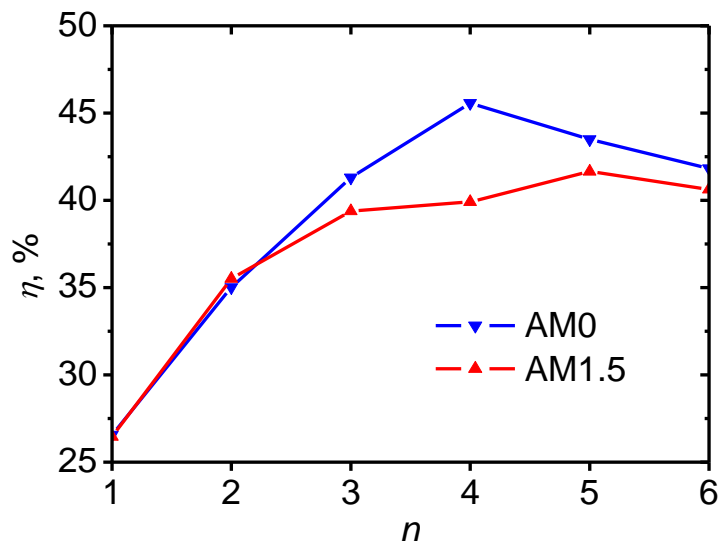

Fig. 10. (Color online) Calculated dependencies of efficiency versus number of cells (the set of experimentally available semiconductors is used) for AM0 and AM1.5 conditions (vertical solar cells).

Finally, Fig. 10 presents theoretical efficiencies $\eta(n)$ for a vertical structure at $d = 10$ μm using parameters of existing semiconductors. (This is in contrast to Fig. 8 and 9, where $\eta$ was calculated for model SE systems). In this realistic case, the maximum efficiencies are comparable to those for a lateral structure, shown in Fig. 6, and the maximums are reached at $n=4$ for AM0 and $n=5$ at AM1.5.

## 4. ANALYSIS OF THE ASSUMPTIONS USED

To simulate dependence of the efficiency $\eta$ of the tandem solar cells on the numbers of subcells $n$ and their temperature $T$ using Eq. (1) with (10)–(14) or (18)–(20), we introduced certain simplifying assumptions. In principle, the above general equations for $\eta(n)$ can be solved without such simplifications. For this purpose, for instance, detailed experimental light absorption coefficients $\alpha(E_g, E_{ph})$ in direct-gap semiconductors with arbitrary band gaps $E_g$ can be used. Since such information is not available for several direct-gap semiconductors, we extrapolate the dependencies from existing results. For the same reason we used in Eq. (10) the same $\nu_i$ for all direct-gap semiconductors for arbitrary $E_g$. We have explained previously why $\nu_i$ are close for a number of reference direct-gap semiconductors, the one used in this work. Furthermore, since $\nu_i \gg 1$, and the photovoltage is proportional to $\ln \nu_i$, our estimates show that the relative error in the calculated efficiency $\eta$ due to this approximation is less than 5%.

In our calculations, we did not take into account band gap variation with temperature. It is known that the smaller the band gap the stronger its variation with temperature (see, *e.g.,* [21]). This effect will be more pronounced under AM0 because of the considerably wider range of SC temperature variation compared to AM1.5 conditions. However, the effect of temperature dependent $E_g(T)$ on the efficiency $\eta(T)$ can be estimated, we illustrate this for the case of GaAs. A decrease of GaAs temperature from 300 K to 264 K under AM0 conditions leads to a less than 1% increase of $E_g$. At the same time, the temperature dependence of the open-circuit voltage $V_s$ (the second term on the right of Eq. (10)), that determines the increase in efficiency, is considerably stronger and results in an efficiency increase by more than 7%. Therefore, neglecting the effect of $E_g(T)$ variation with $T$ is a valid approximation.

## 5. CONCLUSIONS

In photoconversion using tandem solar cells, wide band gap subcells reduce the thermalization losses of high energy photons and small band gap subcells lower the transmission losses of low energy photons, which altogether reduces solar cell (SC) temperature and increases the efficiency $\eta$ in a selfconsistent way. To model the above factors theoretically we combined the detailed balance approach and thermal balance equations. The selfconsistent solution of the general set of equations describes how the efficiency $\eta$ of tandem solar cells depends on the number of subcells $n$, SC temperature $T$, and its heat emission characteristics. Most importantly, we considered the actual photon energy dependence of the light absorption coefficient close to the band edge of semiconductors. The effect of a weak light absorption close to the band edge, generally ignored in the standard approaches (see, *e.g.*, [2,5,12]), leads to the appearance of a maximum in the $\eta(n)$ dependence. Such a maximum has been observed experimentally, but not previously predicted theoretically. Both non-concentrated and concentrated light were considered, and the formalism developed explicitly incorporates the temperature induced modification of $\eta$. In a hypothetical model case with suitable energy gaps of subcells always available and current matching conditions satisfied a shallow maximum of $\eta(n)$ of the tandem SC is achieved at $n =15$ for AM0 and $n =12$ for AM1.5 conditions.

However, when considering technologically available semiconductors with predefined energy gaps $E_g$ and satisfying the current matching requirement a more pronounced maximum of $\eta(n)$ is achieved at lower $n$ (close to 4). As a result, the calculated $\eta(n)$ dependence fits well the shape of the experimentally observed efficiency, the maximum value and its position, as is demonstrated in Fig. 6.

Our selfconsistent simulation of the efficiency dependence on the degree of concentration $\eta(M)$ demonstrates that the optimum performance in the case of concentrated radiation without forced cooling requires a considerable increase of the radiator area to dissipate heat. In the outer space environment, this might be difficult to realize due to weight constraints. However, it follows from our analysis that sufficient SCs temperature reduction and associated efficiency increase under concentrated light conditions can be achieved by improved heat dissipation conditions, realized when both illuminated and shaded SC surfaces radiate as closely as possible to a perfect black body.

To reach a quantitative agreement with experiment, in addition to taking into account radiative recombination, the formalism developed can be further modified to incorporate other limiting factors on the photoconversion efficiency such as, for instance, Shockley-Read and sur-face recombinations [24]. Increase of the surface recombination for tandem SCs with increasing subcells number *n* imposes an extra condition that maximizes *η(n)* even for smaller *n*.

In summary, the theoretical approach proposed offers selfconsistent, more realistic estimates of attainable photoconversion efficiency of tandem SCs as compared to the standard theoretical models available in the literature. The approach and its application demonstrated in the paper can be useful to minimize photocurrent losses in tandem SCs and their optimization.